\documentclass[preprint2]{aastex}

\usepackage{graphicx}
\usepackage{txfonts}
\usepackage{natbib}	
\usepackage{xcolor} 

\slugcomment{Not to appear in Nonlearned J., 45.}

\shorttitle{Collapsed Cores in Globular Clusters}
\shortauthors{Djorgovski et al.}

\begin{document}


\title{Using the Rossiter-McLaughlin effect to observe the transmission spectrum of  Earth's atmosphere \thanks{Based on observations made with the HARPS instrument on the ESO 3.6-m telescope at the La Silla Observatory under the programme ID 093.C-0423(B)} }


\author{F. Yan 
          \altaffilmark{1,2,3},
          R. A. E. Fosbury
		\altaffilmark{3},
          M. G. Petr-Gotzens 
          \altaffilmark{3},
          E. Pall\'e 
          \altaffilmark{4,5}
          and
          G. Zhao
          \altaffilmark{1} }
\email{feiy@nao.cas.cn, gzhao@nao.cas.cn}


\altaffiltext{1}{Key Laboratory of Optical Astronomy, National Astronomical Observatories, Chinese Academy of Sciences, 20A Datun Road, Chaoyang District, 100012 Beijing, China}
\altaffiltext{2}{University of Chinese Academy of Sciences, 19A Yuquan Road, Shijingshan District, 100049 Beijing, China}
\altaffiltext{3}{European Southern Observatory, Karl-Schwarzschild-Str.~2, 85748 Garching bei M\"unchen, Germany}
\altaffiltext{4}{Instituto de Astrof\'isica de Canarias, C/ v\'ia L\'actea, s/n, 38205 La Laguna, Tenerife, Spain}
\altaffiltext{5}{Dpto. de Astrof\'isica, Universidad de La Laguna, 38206 La Laguna, Tenerife, Spain}


\begin{abstract}
Due to stellar rotation, the observed radial velocity of a star varies during the transit of a planet across its surface, a phenomenon known as the Rossiter-McLaughlin (RM) effect. The amplitude of the RM effect is related to the radius of the planet which, because of differential absorption in the planetary atmosphere, depends on wavelength. Therefore, the wavelength-dependent RM effect can be used to probe the planetary atmosphere. We measure for the first time the RM effect of the Earth transiting the Sun using a lunar eclipse observed with the ESO HARPS spectrograph. We analyze the observed RM effect at different wavelengths to obtain the transmission spectrum of the Earth's atmosphere after the correction of the solar limb-darkening and the convective blueshift. The ozone Chappuis band absorption as well as the Rayleigh scattering features are clearly detectable with this technique.
Our observation demonstrates that the RM effect can be an effective technique for exoplanet atmosphere characterization. Its particular asset is that photometric reference stars are not required, circumventing the principal challenge for transmission spectroscopy studies of exoplanet atmospheres using large ground-based telescopes.
\end{abstract}






\section{Introduction}
Among the nearly 2000 exoplanets discovered so far, more than half are transiting systems. Due to the rotation of the host star, the observed stellar radial velocity (RV) is expected to change as the planet transits different parts of the rotating stellar surface. This is called the Rossiter-McLaughlin (RM) effect and was initially measured for eclipsing binary stars \citep{Rossiter1924,McLaughlin1924}. After the first observation of the exoplanetary RM effect by \cite{Queloz2000}, there are now more than 80 exoplanet systems with this effect observed, e.~g. \cite{Albrecht2012}. This effect is normally used to measure the projected angle between the stellar spin and the planetary orbit. 

\cite{Snellen2004} was the first to propose using the wavelength-dependent RM effect to probe the atmospheres of transiting exoplanets. This method exploits the fact that the planetary effective radius is wavelength-dependent due to differential atmospheric absorptions.
\cite{Dreizler2009} theoretically modeled the RM effect for the Na absorption of giant exoplanets.
The advantage of the RM method compared to the traditional spectrophotometry is that it does not demand  the use of photometric reference stars. The RM method promises to become a powerful technique for future transmission spectrum measurements, especially as the next generation of very large ground-based telescopes are likely to have a small field-of-view, making it difficult to employ suitable reference stars. 

With more and more terrestrial exoplanets being discovered, the characterization of their atmospheres will become a major goal for exoplanet research in which the RM method can play an important role. 
Although it is currently difficult to observe the RM effect of these terrestrial exoplanets due to instrumental limitations, a lunar eclipse provides us with an opportunity to observe the Earth transiting the Sun and so explore the effectiveness of the RM method. The concept of regarding the Earth as an exo-Earth using lunar eclipses has been applied in the past by \cite{Palle2009,Vidal2010,Ugolnikov2013,Arnold2014,Yan2015b}. These studies obtained the transmission spectrum directly from the ratio spectrum before and during a lunar eclipse.
Here we present for the first time the use of the RM effect of an Earth transit during a lunar eclipse to retrieve the transmission spectrum.

\section{Observations and RM effect measurement}
\begin{figure*}
   \centering
   \includegraphics[width=0.50\textwidth]{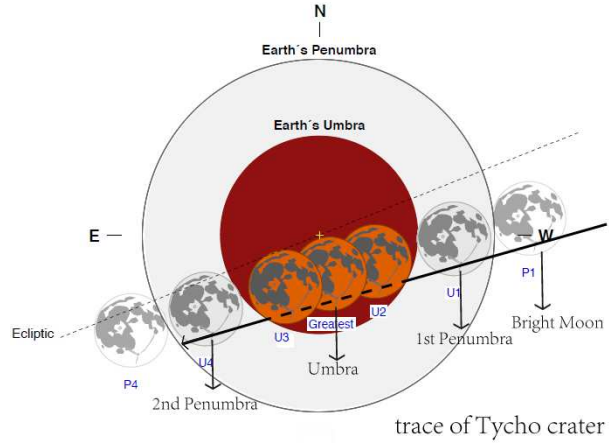}
      \caption{ Trajectory of the observed position (Tycho crater). Our observation began at the bright Moon stage and lasted until the 2nd penumbra.
      The figure is reproduced from the NASA lunar eclipse page.
      }
         \label{NASA-Tycho}
   \end{figure*}

We observed the 15-April-2014 lunar eclipse with the High Accuracy Radial velocity Planet Searcher (HARPS) mounted on the ESO La Silla 3.6m telescope \citep{Mayor2003}. 
A consecutive sequence of observations over one entire night covered all of the eclipse stages, i.e. the penumbral and umbral eclipse and out of eclipse (called hereafter the bright Moon). 
The classical fiber spectroscopy mode was used with the fiber located at the center of the Tycho crater (see Fig.~\ref{NASA-Tycho} for the trajectory). 
The telescope used non-sidereal tracking and the tracking velocities were updated every few minutes. In addition, manual guiding on crater structures was performed. This worked reasonably well and limited the pointing coordinate drift to $\lesssim$ 1.5 arcmin.
We used varying exposure times during the eclipse because the lunar surface brightness changed dramatically.
In total, 382 lunar spectra were obtained (Table \ref{observation}). 

%
\begin{table*}
\caption{Details of the observed spectra.}             
\label{observation}      
\centering                          
\begin{tabular}{c c c c c}        
\hline\hline                 
Eclipse stage	&	UT time span	 &	Number of observed spectra	&	Exposure time (second)	&	Airmass range\\     
\hline                       
 		Bright Moon	& 00:30 - 05:32	&	294	&	30	& 1.98 $\sim$ 1.09\\
        1st Penumbra	& 05:32 - 06:45	&	55 	&	30,100,500	& 1.09 $\sim$ 1.23	\\
        Umbra	& 06:45 - 08:47 &	13 &	500	& 1.23 $\sim$ 2.01	\\
        2nd Penumbra	& 08:47 - 09:35 &	20	&	500 $\sim$ 60	& 2.01 $\sim$ 2.90\\
\hline                                   
\end{tabular}
\end{table*}

\begin{figure*}
   \centering
   \includegraphics[height=0.4\textwidth]{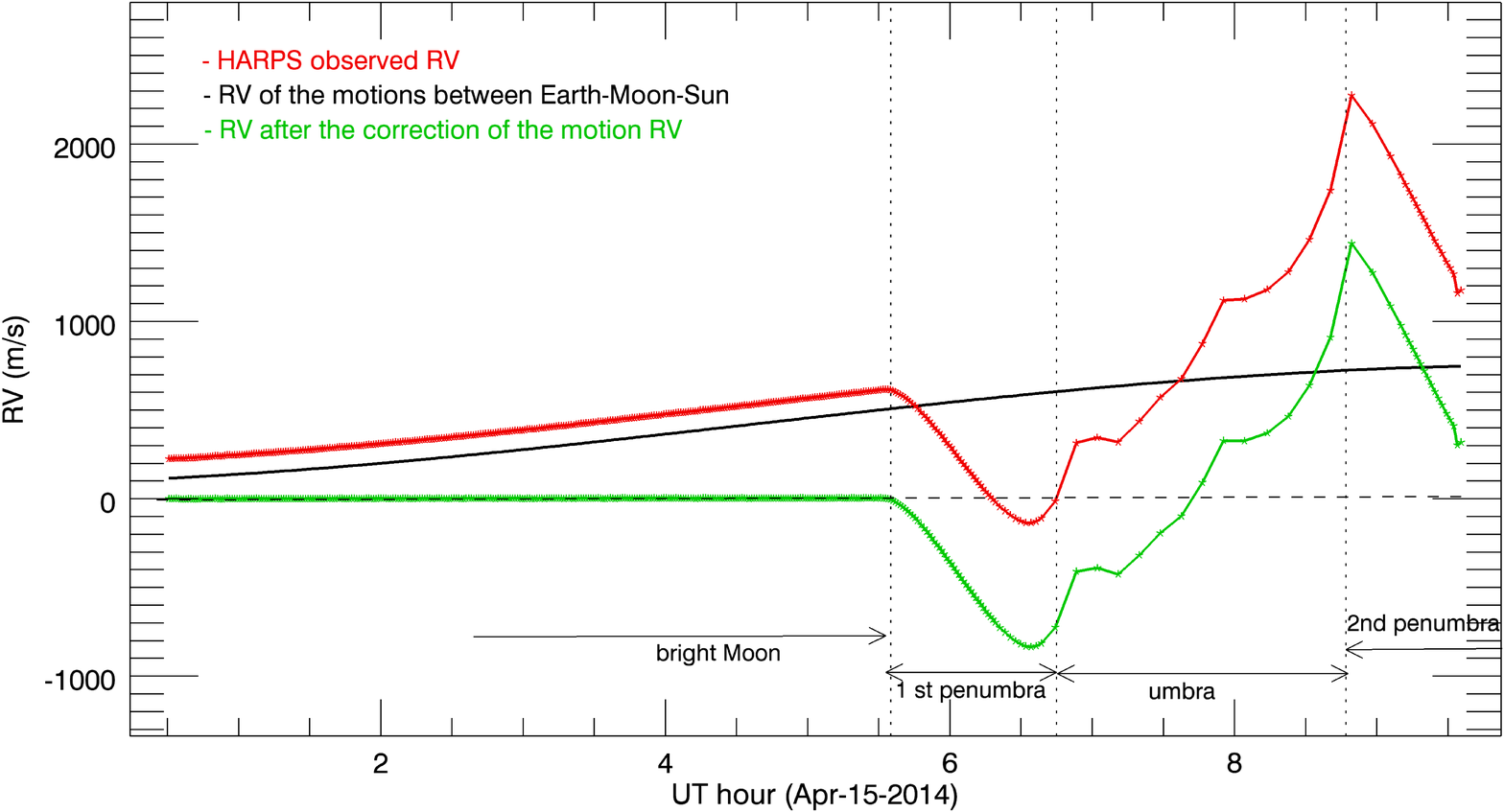}
      \caption{The observed RV (red line) and the theoretical RV of the motions between the Earth, the Moon and the Sun (black line).
      The observed RV after the corrections of the motions and the baseline offset (green line), which is the RM curve of the Earth transiting the Sun. 
      Positive values represent redshift. 
      }
         \label{RME-curve}
   \end{figure*}

Data reduction was performed using the HARPS pipeline. Each exposure frame comprises 72 spectral orders that cover the wavelength range from 378 nm to 691 nm with a spectral resolution of $\lambda / \Delta \lambda$ $\sim$ 115,000. Radial velocities are measured, for each spectral order as well as for the overall spectral range, using a cross-correlation with a G2 stellar spectral template.  The overall measured and theoretical RV curves are shown in Fig.~\ref{RME-curve}. The theoretical RV is the combination of two components: the motion of the Sun with respect to Tycho and the motion of Tycho with respect to the observer. These velocities are calculated with the JPL Horizon Ephemeris\footnote{Http://ssd.jpl.nasa.gov/?ephemerides}, which considers the orbital motions, rotations of the Earth and the Moon, and the light travel time corrections.

There is an offset between the theoretical RV and the measured RV, which probably originates from the RV zero-point of the spectral template used by the HARPS pipeline \citep{Molaro2013}. We correct this offset by using the data points taken during the bright Moon as a baseline. 
Fig.~\ref{RME-curve} shows the measured RV with the motion RV corrected and the baseline offset subtracted. This is the final RV curve of the RM effect for the Earth transiting the Sun.
At the bright Moon stage, the RV is essentially corrected to zero, but there is still a small slope with an amplitude of about 4 m/s which is probably due to the instrumental drift and the non-perfect telescope guiding when observing the Moon. Since the RM amplitude is on the order of 2 km/s, this small residual slope does not significantly affect our analysis.

When the Moon enters the 1st penumbra, the observed RV becomes negative (blueshift) as the Earth begins to transit the redshifted rotating part of the solar disk. At the umbral stage, the RV gradually changes from a blueshift to a redshift. 
When the Moon enters the 2nd penumbral stage, the RV is redshifted since the Earth obscures mainly the blueshifted  solar region. The RV gradually decreases during this stage as the Moon moves out of the penumbral shadow.

The umbral RV has a relatively complex structure and its details are determined by a combination of both the refracted part of the solar disk and the properties of the Earth's atmosphere refracting the sunlight. 
In this work, the umbral part is not studied since we focus only on the penumbral parts for the RM effect.


\section{Retrieving the transmission spectrum}

\subsection{Modelling the RM effect}

   \begin{figure*}
   \centering
   \includegraphics[width=0.80\textwidth, height=0.45\textwidth]{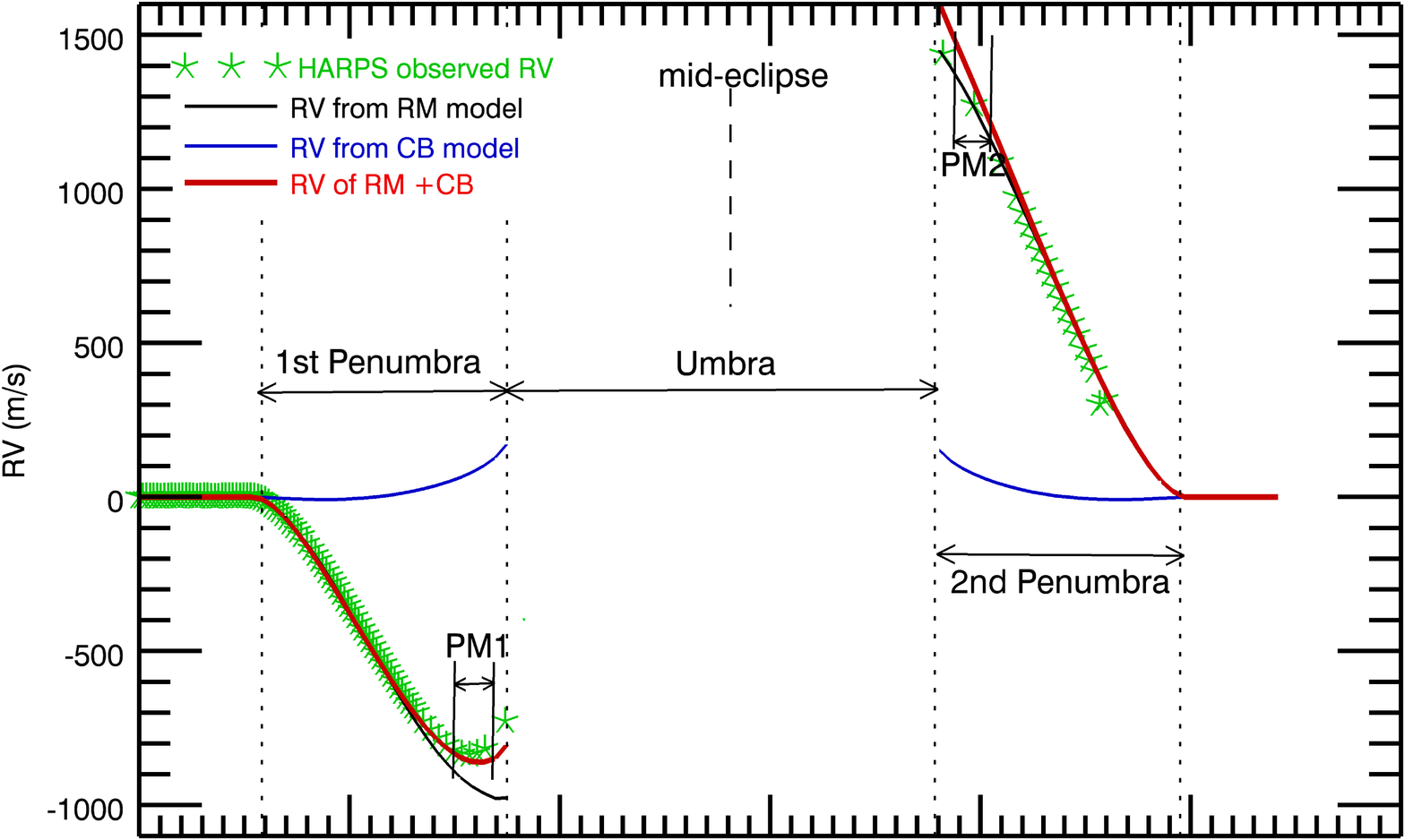}
      \includegraphics[width=0.777\textwidth, height=0.10\textwidth]{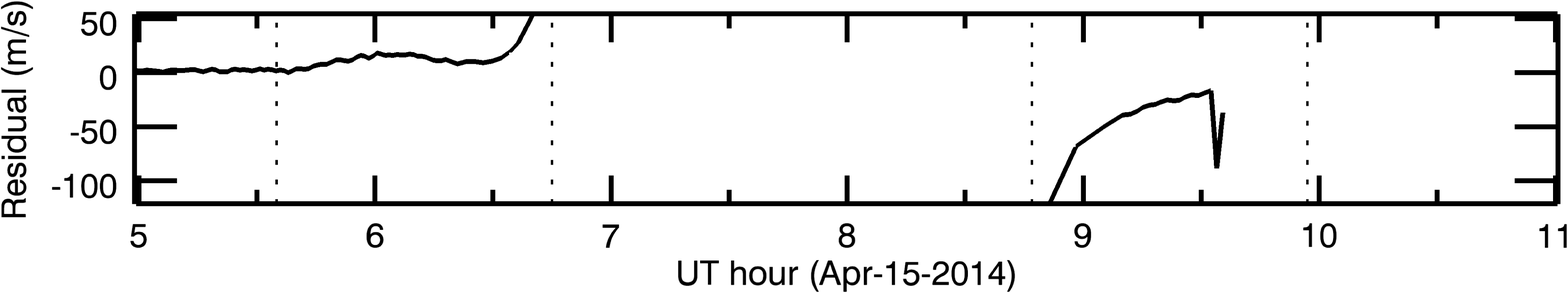}
      \caption{The observed RM effect curve (green points) together with the modelled RV curve (red line), which is the combination 
 of the white-light RM model (black line) and the convective blueshift model (blue line).
See the text for the details of the parameters used in the model. 
Two regions on the RV curve which are symmetric with respect to the mid-eclipse (labelled as PM1 and PM2) are chosen to calculate the RM amplitude.  
PM1 contains 4 observed points with a time span of 496 s and PM2 contains one observed point with a time span of 500 s.
The lower panel shows the residual between the observed and the modelled RV.
}
         \label{white-light-model}
   \end{figure*}

To retrieve the wavelength-dependent RM effect due to the Earth's atmospheric absorption, we firstly establish an RM model
to take into account the wavelength-dependent solar parameters. 

Our model follows the method described in \cite{Gaudi2007} and \cite{Haswell2010}. We divide the solar disk into elements with a size of $0.01\mathrm{R_\odot} \times0.01\mathrm{R_\odot}$ and the radial velocity due to solar rotation for each element is:
\begin{equation}
      v(x,y) = \omega~x~\mathrm{R_\odot}~\mathrm{sin}~I_s
   \end{equation}
where $\omega$ is the rotation angular velocity, $I_s$ is the inclination of the solar spin axis towards Tycho, x and y are in the sky-plane coordinate system of which the origin is at the projected solar center and the y-axis is along the projected solar rotation axis. The coordinate values are in units of the solar radius $\mathrm{R_\odot}$.
The RV anomaly $\bigtriangleup V(t)$ due to the RM effect is then calculated by integrating the intensity-weighted RVs of the visible solar disk:
\begin{equation}
\bigtriangleup V(t) = \frac{\int \int v(x,y)~I(x,y)~\mathrm{d}x~\mathrm{d}y}{\int \int I(x,y)~\mathrm{d}x~\mathrm{d}y} .
   \label{equ-deltaRV}   
   \end{equation}
The RM curve for white-light (i.e. the entire wavelength range of the HARPS spectrum) is generated from this model (the black line in Fig.~\ref{white-light-model}). The parameters are further described as follows:
   \begin{enumerate}	
      \item the coordinates of the Earth's center and the solar center as seen from Tycho are generated using the Horizon Ephemeris. 
      We use the Earth radius of $\mathrm{R_\oplus}=6378$ km and the solar radius of  $\mathrm{R_\odot}=6.955\times10^5$ km. These data are used to determine the visible part of the solar disk at a given time.
      \item the inclination $I_s$ is calculated using the solar north pole position angle given by Horizon. The mean value of $I_s$ is $95^{\circ}.61$ and changes by about $0^{\circ}.02$ during the eclipse.
      \item the solar differential rotation used is \citep{Cox2000}:
\begin{equation}
      \omega=A + B \mathrm{sin}^2\phi \, \mathrm{(deg/day)} ,
   \end{equation}      
where $\phi$ is the heliographic latitude and is calculated with:
\begin{equation}
      \mathrm{sin}~\phi = (1-x^2-y^2)^{1/2}~\mathrm{cos}~I_s+y~\mathrm{sin}~I_s.
   \end{equation}
   The coefficients adopted are A = 13.46, B = -2.99 \citep{Dupree1972}.
	\item the quadratic limb-darkening coefficients \citep{Claret2004} are used for the white-light model to calculate the intensity $I(x,y)$. We adopt the coefficients used by \cite{Molaro2013}, i.e. $u_a = 0.5524$ and $u_b=0.3637$.
   \end{enumerate}

For an exoplanet transit, the projected angle between the stellar spin and the planetary orbit is calculated by fitting the observed RV with the RM model. However, we adopt the actual data from the Horizon Ephemeris and so the model curve in Fig.~\ref{white-light-model} is not the result of such a fit. The ``projected spin-orbit" angle for this observation is $-169^{\circ}.7$ calculated using the trajectory of the Earth as seen from Tycho. Because this trajectory is the combination of the Earth and lunar orbits, the angle is not the real spin-orbit angle of the Sun-Earth system of $7^{\circ}.25$.

\subsection{Other effects}
\subsubsection{Convective blueshift}
The stellar convective motion results in a blueshift of the lines which is called the convective blueshift (CB) \citep{Dravin1982}.
As the CB radial component varies across the stellar disk, the observed stellar RV changes when a planet transits the stellar surface. \cite{Shporer2011} discusses the CB during an exoplanet transit and its effect on the RM effect. 
We follow the CB model in \cite{Shporer2011} and adopt a typical CB value of the Sun $V_\mathrm{CB} \approx -300$ m/s for the white-light RV calculation. 
The blueshift value for each element on the solar disk is $\mu V_\mathrm{CB}$, where $\mu = (1-x^2-y^2)^{1/2}$. 
The RV change caused by CB (blue line in Fig.~\ref{white-light-model}) is then calculated in a similar way to the RM model and added to the RV value caused by the RM effect to give the red line in Fig.~\ref{white-light-model}.

\subsubsection{Atmospheric refraction}
The effects of atmospheric refraction are important for interpreting the transits of terrestrial planets \citep{Garcia2012,Betremieux2014} and
affect the RM effect. As the atmosphere refracts the light from the stellar disk region that should be obscured by the planet, the apparent effective radius of the planet will appear smaller than without refraction. Thus in general, refraction would result in a smaller RM amplitude than without refraction. 
In our RM model, we do not include atmospheric refraction, which means the modelled RM amplitude should be larger than the actual observed RM amplitude. This is consistent with the result in Fig.~\ref{white-light-model} in which the modelled RV value without refraction (red line) is generally larger than the observed RV (green points). 

\subsubsection{Line broadening mechanisms}
\label{sec-turbulence}
In the RM model described above, we assume the line profile is dominated by the stellar rotation and use the first-moment approximation \citep{Gaudi2007, Ohta2005}. This assumption gives us the RV anomaly expressed by Equation \ref{equ-deltaRV}. However, other mechanisms like the microturbulence, macroturbulence and instrumental broadening also affect the line profile and can change the shape of the RV curve \citep{Hirano2011}. This may contribute to the discrepancy between the modelled RV and the observed RV in Fig.~\ref{white-light-model}.

\subsection{The RM effect in different wavelength ranges}
The HARPS pipeline produces directly a RV value for each of the 72 spectral orders. Three orders do not have useful RV data due to the lack of stellar lines in the corresponding spectral mask. After correcting for the Earth-Moon-Sun motion and the baseline offset described in Section 2, we obtain 69 RM curves each representing a different wavelength range.

   \begin{figure*}
   \centering
   \includegraphics[height=0.3\textwidth, width=0.5\textwidth]{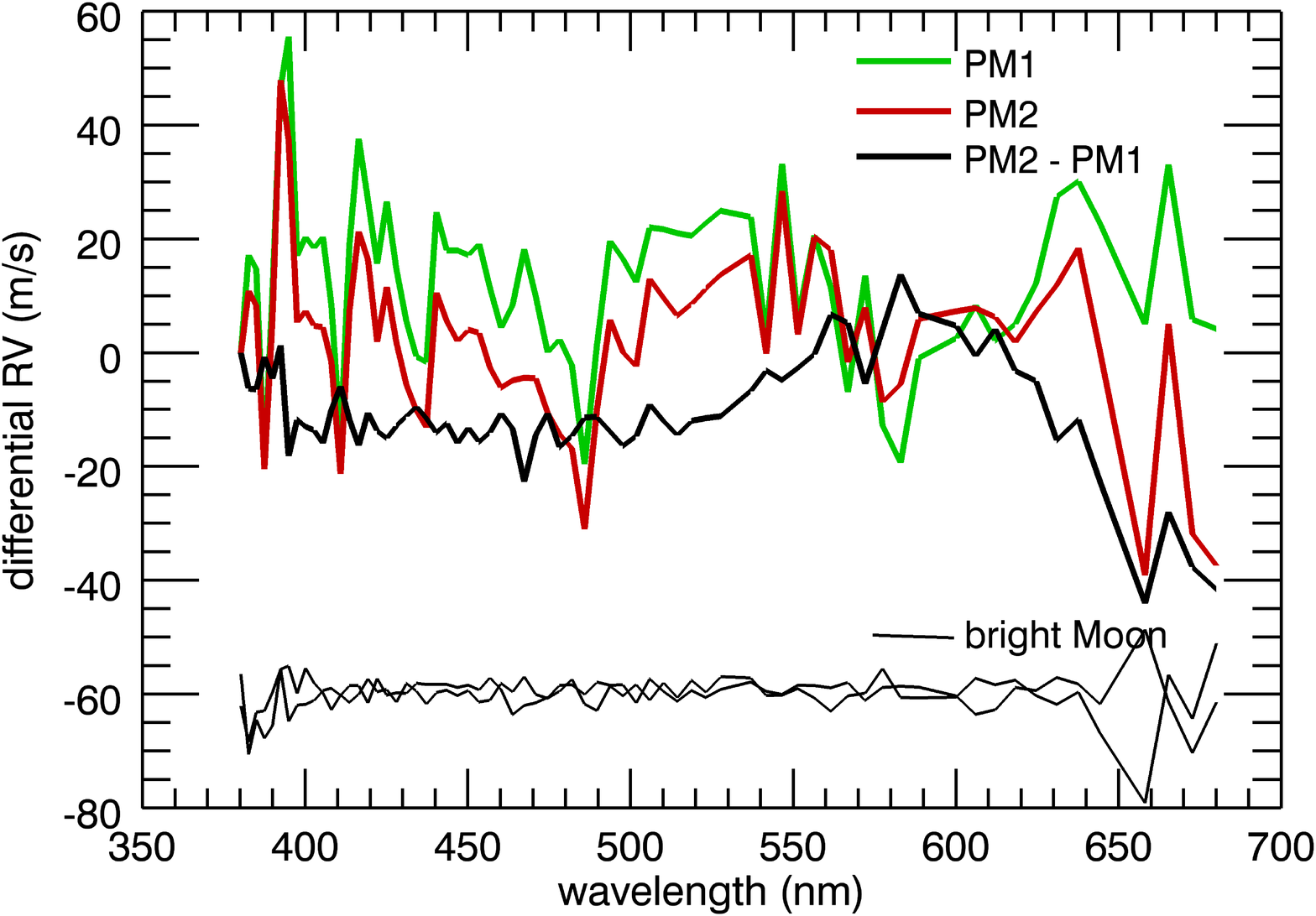}
   \includegraphics[height=0.3\textwidth, width=0.5\textwidth]{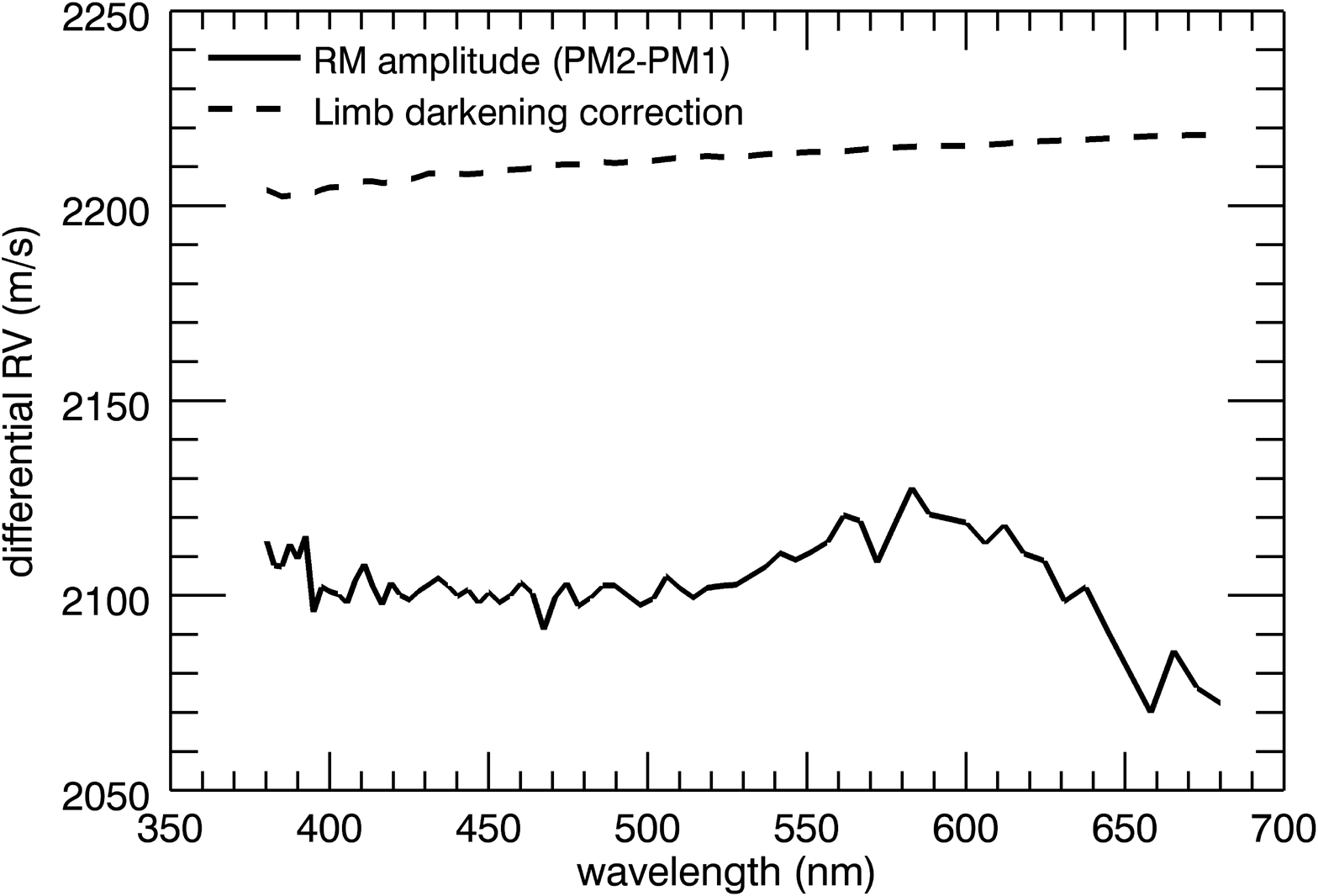}
   \includegraphics[height=0.3\textwidth, width=0.5\textwidth]{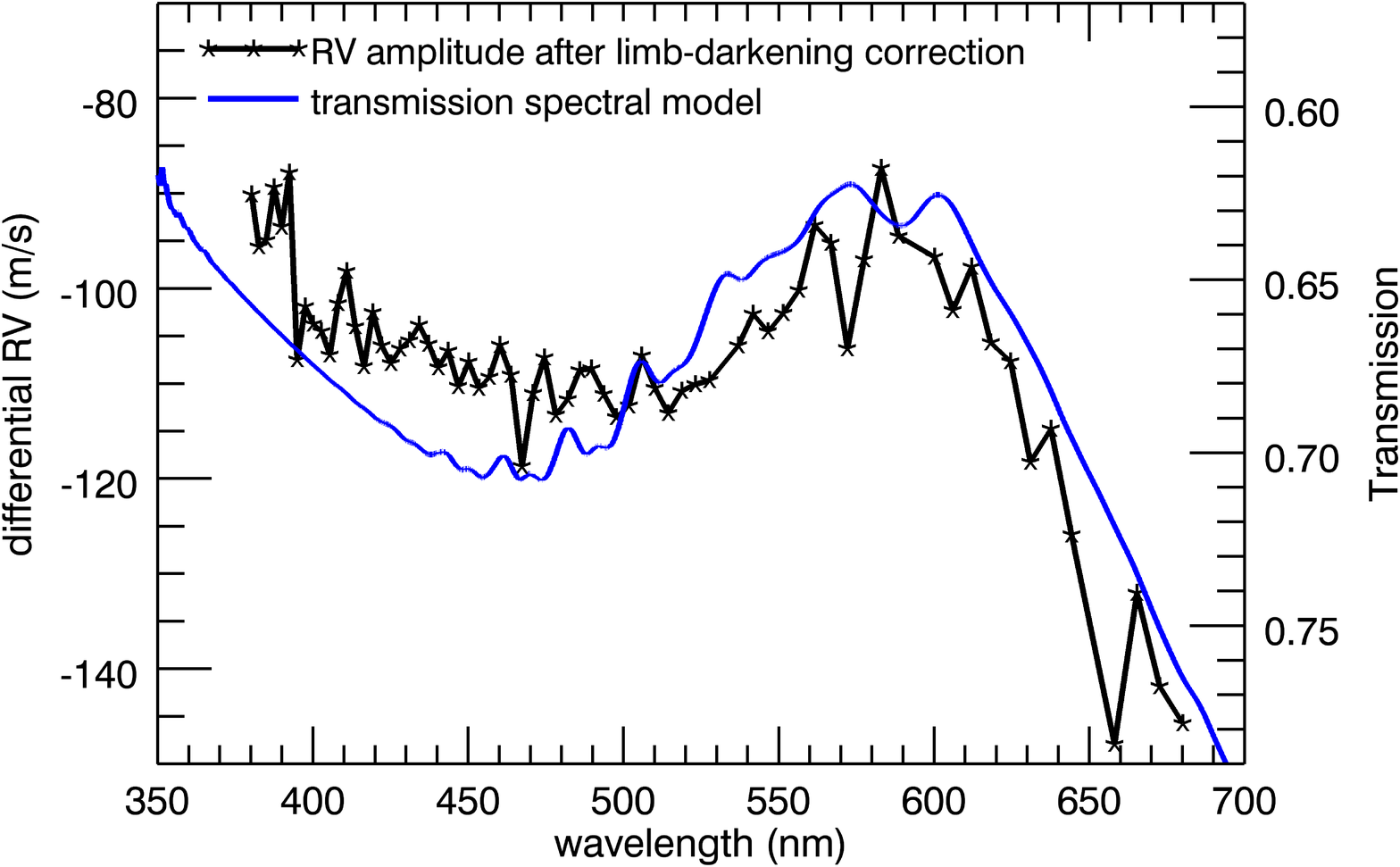}
      \caption{\textbf{(a)} Top panel: the RV values of PM1 (green line) and PM2 (red line) for the 69 spectral orders. The black line is the value of PM2 - PM1. All three lines here are shifted to zero at the first point (380 nm) in order to compare their structures. 
 At the bottom, the RV curves of two bright Moon spectra are plotted (their values are shifted to - 60 m/s), and they are relatively flat compared to PM1 and PM2. The order at 658 nm has a low RV accuracy due to a low number of available solar lines.
      \textbf{(b)} Middle panel: model values of PM2 - PM1 considering the wavelength-dependent limb-darkening coefficients with a fixed Earth radius (dashed line). The black line is the observed PM2 - PM1 as in the top panel.
      \textbf{(c)} Bottom panel: the difference between the observed and modelled PM2 - PM1 values shown in (b). This is the final RM amplitude in different wavelength ranges after the limb-darkening correction. The HARPS orders' central wavelengths are indicated by asterisks.
      The transmission spectral model of the Earth's atmosphere (blue line) is plotted for comparison. 
      }
         \label{RME-amplitude}
   \end{figure*}

To obtain the RM effect differences caused by the Earth's atmosphere, the wavelength-dependent solar parameters need to be considered.
The first aspect concerns the stellar convection blueshift. 
The actual RV caused by CB depends on where the line forms, e.g. the strong lines or low-excitation lines may have a small CB because they usually form high in the stellar atmosphere where the granulation is weak \citep{Dravin1982,Gray2009}. 
As each of the HARPS spectral orders comprises stellar lines with different CB values, the $V_\mathrm{CB}$ of each spectral order varies.
Therefore, unlike the white-light CB model for which we can assume $V_\mathrm{CB} \approx -300$ m/s, the actual CB effect for a given HARPS order is difficult to model.
Instead, we use two symmetric parts on the RV curve to cancel this effect utilizing the fact that the convection induced RV anomaly during transit is symmetric with respect to the mid-transit as shown by the blue line in Fig.~\ref{white-light-model}.
Two regions of the RM curve (labelled as PM1 and PM2 in Fig.~\ref{white-light-model}) are then chosen. The PM1 and PM2 are symmetric regions of the eclipse, i.e.\ they are at the same distance from mid-eclipse.
Fig.~\ref{RME-amplitude}a shows the RV values of PM1 and PM2 and the RV difference between them (i.e. PM2 - PM1). We use this PM2 - PM1 value to represent the RM effect amplitude for each spectral order.
From Fig.~\ref{RME-amplitude}a, it can be seen that there is a correlated systematic difference between PM1 and PM2. 
We believe that this correlated pattern results from the measured effect of the CB depending on the type and number of spectral lines present in each spectral order but is well-cancelled by using PM2 - PM1.

The second wavelength-dependent solar parameter is the limb-darkening. Here we use the empirical power-law limb-darkening coefficients from \cite{Hestroffer1998}. For each of the HARPS orders, we interpolate a limb-darkening coefficient at the corresponding wavelength and calculate a value of PM2 - PM1 with a fixed Earth radius (6378 km) using our RM model. The model values of PM2 - PM1 are shown in Fig.~\ref{RME-amplitude}b. The final RM effect at different wavelength ranges, after the correction of the limb-darkening, is presented in Fig.~\ref{RME-amplitude}c.
The limb-darkening correction made here is mainly for the continuum, however, since the solar lines have different limb-darkening compared to the adjacent continuum \citep{Snellen2004, Yan2015a}, this could introduce some extra noise in the final RM amplitudes.

\subsection{Results and discussion} 
The RV curve of the wavelength-dependent RM amplitudes in Fig.\ref{RME-amplitude}c results from the different effective radius at different wavelengths, and it can be regarded as a mapping of the low resolution transmission spectrum of the Earth's atmosphere into effective radius and hence radial velocity space.
To interpret the observed atmospheric features, we build a transmission spectral model following the methods of \cite{Kaltenegger2009} and \cite{Yan2015b}. 
We calculate the overall transmission spectrum of the Earth atmosphere from 0 to 80 km altitude considering
the ozone absorption and Rayleigh scattering extinction (blue line in Fig.~\ref{RME-amplitude}c). This transmission spectrum is overlaid with the RM amplitude curve to compare their shapes. We emphasize here that the transmission spectrum model is not used to fit the observed RM amplitude curve but to demonstrate the presence of the atmospheric features mapped into it. By comparing the spectral shapes in Fig.~\ref{RME-amplitude}c, we interpret the RM amplitude curve as follows: the RM amplitude is larger towards the blue due to Rayleigh scattering extinction that makes the atmosphere more opaque and the atmospheric effective thickness larger at shorter wavelengths. The broad peak around 600 nm results from the ozone Chappuis band absorption while the RM amplitude towards the red becomes smaller since both the ozone absorption and the Rayleigh scattering become weaker, rendering the atmosphere more transparent.

The RM model we have used for this work suffers from several defects and incompleteness. The first is observational in that there is a slight, but barely significant, drift in the telescope guiding on the lunar surface. 
The second is that other line broadening mechanisms like the stellar micro/macro-turbulence are not included in the RM model (cf. subsection \ref{sec-turbulence}). 
The third is more fundamental and its full solution is beyond the scope of this letter. This is the determination of the detailed mapping between the RM amplitude and the atmospheric exctinction as a function of wavelength. This depends on the effects of atmospheric refraction in the exoplanetary atmosphere which will be influenced not only by the absorption coefficients of its gaseous constituents but also by the effects of screening due to clouds and aerosols.

The combination of these effects can explain the slight deviation of the modelled RV from the observed RV values. 
It can also explain why the modelled RM amplitudes for the limb darkening correction are larger than the observed RM amplitudes (shown in Fig.~\ref{RME-amplitude}b). However, this has a very limited effect on our retrieval of the final RM amplitude curve since the limb-darkening correction produces only a small RV variation with wavelength so that the shape of the final RM amplitude curve is little affected by the RM model. Future modelling work containing a proper treatment of refraction and stellar turbulence will allow us to combine the transmission spectral model with the RM model, and to fit the observed RM amplitude curve directly instead of just comparing its shape with the transmission spectrum as presented in the current work.

\section{Conclusion}
We have observed for the first time the Rossiter-McLaughlin effect of the Earth transiting the Sun using a lunar eclipse. The RM effect curve has been obtained using high accuracy RV observations and an RM effect model is built to analyze the observed result.
Separate RM curves at different wavelengths are obtained from 69 HARPS spectral orders.
After the correction of the wavelength-dependent limb-darkening of the Sun and the convective blue-shift of the solar lines, we retrieve the wavelength-dependent RM amplitudes due to the transmission of the Earth's atmosphere. The ozone Chappuis band and the Rayleigh scattering signatures are clearly detectable.

The RM method can be used to detect broad features, such as the Rayleigh scattering, in exoplanet atmospheres. The advantage is that no reference stars are needed in contrast to the requirements for the traditional spectrophotometric method.
Since the next-generation ground-based telescopes such as E-ELT will have a relatively small field-of-view, limiting the access to nearby reference stars,
the RM method will provide a promising technique for the characterization of planetary atmospheres.
Thus, in the future, this method can be applied to detect the atmospheres of terrestrial planets and particularly to search for the bio-signature gas ozone. 

Extending the observation to near-infrared (NIR) wavelengths will yield more absorption features. However, the ability of the RM method is limited by the number of suitable spectral lines of the parent star. The Sun, for example, does not have sufficient lines in the NIR. 
However, for M-type stars which are expected to be promising targets for exoplanet atmosphere studies, there are many stellar molecular lines in the NIR which can be exploited.





\acknowledgments
We greatly appreciate the excellent support from the LaSilla/Paranal Science Operations team. In particular, we thank Valentin Ivanov for performing the observations on site and his and Lorenzo Monaco's support during the preparation of the observing run.
We would also like to thank Gaspare Lo Curto and Andrea Chiavassa for helpful discussions and the referee for useful suggestions. 
The study is supported by the National Natural Science Foundation of China under grants Nos. 11390371 and 11233004.



{\it Facilities:} \facility{ESO-3.6m (HARPS)}.




%

\clearpage

\end{document}